\documentclass[runningheads]{llncs}
\usepackage[nonumberlist,nomain,nopostdot,acronym,toc,nogroupskip,symbols]{glossaries}
\usepackage{todonotes}
\usepackage{url}
\usepackage{xurl}
\usepackage[breaklinks]{hyperref}
\usepackage{cleveref}
\usepackage{graphicx}
\usepackage{tabularx}
\usepackage{booktabs}
\usepackage{paralist}
\usepackage{placeins}
\usepackage{microtype}
\usepackage{amsfonts} 
\usepackage{pifont}

\newcommand{\cmark}{\ding{51}}%
\newcommand{\xmark}{\ding{55}}%
\newcommand{\mypar}[1]{\smallskip\noindent\textbf{#1.}}

\usepackage[acronym]{glossaries}
\newacronym[\glslongpluralkey={Business Processes}]{bp}{BP}{Business Process}
\newacronym{bi}{BI}{Business Intelligence}
\newacronym{bpi}{BPI}{Business Process Intelligence}
\newacronym{bpm}{BPM}{Business Process Management}
\newacronym{bpmn}{BPMN}{Business Process Model and Notation}
\newacronym{bpms}{BPMS}{Business Process Management System}
\newacronym{copd}{COPD}{Chronic Obstructive Pulmonary Disease}
\newacronym{crm}{CRM}{Customer Relationship Management}
\newacronym{dbms}{DBMS}{Database Management System}
\newacronym{dsr}{DSR}{Design Science Research}
\newacronym{dt}{DT}{Decision Tree}
\newacronym{ecg}{ECG}{Event-Coupling Graph}
\newacronym{epc}{EPC}{Event-Driven Process Chain} 
\newacronym{er}{ER}{Entity-Relationship}
\newacronym{etl}{ETL}{Extract, Transform and Load}
\newacronym{feds}{FEDS}{Framework for Evaluation in Design Science Research}
\newacronym{floss}{FLOSS}{Free Libre and Open Source Software} 
\newacronym{fmc}{FMC}{Fundamental Modeling Concepts}
\newacronym{gt}{GT}{Grounded Theory}
\newacronym{gui}{GUI}{Graphical User Interface}
\newacronym{ide}{IDE}{Integrated Development Environment}
\newacronym{it}{IT}{Information Technology}
\newacronym{its}{ITS}{Issue Tracking System}
\newacronym{kpi}{KPI}{Key Performance Indicator}
\newacronym{obda}{OBDA}{Ontology-based Data Access}
\newacronym{oss}{OSS}{Open-source Software}
\newacronym{lca}{LCA}{Least Common Ancestor}
\newacronym{loc}{LOC}{Lines of Code}
\newacronym{owl}{OWL}{Web Ontology Language}
\newacronym{msr}{MSR}{Mining Software Repositories}
\newacronym{nlp}{NLP}{Natural Language Processing}
\newacronym{pais}{PAIS}{Process-Aware Information Systems} 
\newacronym{pert}{PERT}{Program Evaluation Review Technique} 
\newacronym{pm}{PM}{Process mining}
\newacronym{pn}{PN}{Petri net}
\newacronym{ppi}{PPI}{Process Performance Indicator}
\newacronym{pr}{PR}{Pull Request}
\newacronym{ram}{RAM}{Reliability, Availability and Maintainability}
\newacronym{rbac}{RBAC}{Role-Based Access Control}
\newacronym{rup}{RUP}{Rational Unified Process}
\newacronym{scm}{SCM}{Software Configuration Management} 
\newacronym{sme}{SME}{Small-Medium Enterprise}
\newacronym[\glslongpluralkey={Software Development Methodologies}]{sdm}{SDM}{Software Development Methodology}
\newacronym{soa}{SOA}{Service-Oriented Architecture}
\newacronym{svn}{SVN}{Subversion}
\newacronym{tm}{TM}{Text Mining}
\newacronym{xes}{XES}{eXtensible Event Stream}
\newacronym{vcs}{VCS}{Version Control System}
\newacronym{wf}{WF}{Workflow}
\newacronym{wfms}{WfMS}{Workflow Management System}
\newacronym{pw}{PW}{Project Workload}
\newacronym{tw}{TW}{Type of change Workload}
\newacronym{nap}{NAP}{Number of Authors in Project}
\newacronym{ntp}{NTP}{Number of Types of work in Project}
\newacronym{pis}{PIS}{Specialization of Author in each Activity Type}
\newacronym{rpis}{RPIS}{Specialization of Relative Author in each Activity Type}
\newacronym{pws}{PWS}{Specialization of Project Workload}
\newacronym{rpws}{RPWS}{Specialization of Relative Project Workload}
\newacronym{uti}{UTI}{User Activity Involvement}
\newacronym{utw}{UTW}{User Activity Workload}
\newacronym{who}{WHO}{Wold Health Organization}

\newacronym{yawl}{YAWL}{Yet Another Workflow Language}

\newglossaryentry{alpha}{name={$\alpha$},description={the $\alpha$ imperative process discovery algorithm}}
\newglossaryentry{alphap}{name={$\alpha+$},description={the $\alpha+$ imperative process discovery algorithm}}
\newglossaryentry{alphapp}{name={$\alpha++$},description={the $\alpha++$ imperative process discovery algorithm}}
\newglossaryentry{minerful}{name={MINERful},description={the MINERful declarative process discovery algorithm}}
\newglossaryentry{heuri}{name={Heuristic},description={the Heuristic Miner imperative process discovery algorithm}}
\newglossaryentry{declare}{name={\textsc{Declare}},description={the \textsc{Declare} declarative process modelling language}}
\def\LogAlph {\ensuremath{\mathcal{A}}}
\newglossaryentry{logalph}{
	name={log alphabet},description={process alphabet, as reflected in the log},%
	symbol={\LogAlph}}

\newglossaryentry{evtuniv}{
	name={log alphabet},description={universe of events},%
	symbol={\LogAlph}}

\usepackage[framemethod=TikZ]{mdframed}

\begin{document}
\title{Unraveling the Never-Ending Story of \\Lifecycles and Vitalizing Processes}
\titlerunning{Unraveling the Never-Ending Story of \\Lifecycles and Vitalizing Processes}
\authorrunning{Fahrenkrog-Petersen et al.}

\author{Stephan A. Fahrenkrog-Petersen\inst{1,2}\orcidID{0000-0002-1863-8390} \and
Saimir Bala\inst{2}\orcidID{0000-0001-7179-1901} \and
Luise Pufahl\inst{3}\orcidID{0000-0002-5182-2587}\and
Jan Mendling\inst{1,2,4}\orcidID{0000-0002-7260-524X}}

\institute{Weizenbaum Institute for the Networked Society, Berlin, Germany \and
Humboldt-Universität zu Berlin, Berlin, Germany
\email{\{stephan.fahrenkrog-petersen,saimir.bala,jan.mendling\}@hu-berlin.de}\\ \and
Technical University of Munich, Heilbronn, Germany
\email{luise.pufahl@tum.de}\\ \and
Wirtschaftsuniversität Wien, Vienna, Austria }

\maketitle              %
\begin{abstract}
Business process management (BPM) has been widely used to discover, model, analyze, and optimize organizational processes. BPM looks at these processes with analysis techniques that assume a clearly defined start and end. However, not all processes adhere to this logic, with the consequence that their behavior cannot be appropriately captured by BPM analysis techniques.
This paper addresses this research problem at a conceptual level. More specifically, we introduce the notion of vitalizing business processes that target the lifecycle process of one or more entities. We show the existence of lifecycle processes in many industries and that their appropriate conceptualizations pave the way for the definition of suitable modeling and analysis techniques. This paper provides a set of requirements for their analysis, and a conceptualization of lifecycle and vitalizing processes.

\end{abstract}

\section{Introduction}
\label{sec:intro}
Business process management~\cite{dumas2013fundamentals} (BPM) is concerned with the discovery, modeling, analysis and optimization of business processes. These processes structure the different operational tasks within an organization, for instance, the hiring of a new employee or the purchasing of some supplies.
BPM provides technological and methodological support for organizations to improve business processes.
To this end, a wide range of methods and corresponding tools are available, such as data-driven analysis with the help of process mining~\cite{van2012process} or automation of sub-routines by use of robotic process automation~\cite{van2018robotic}. 

So far, the BPM literature has mainly focused on business processes that are directed towards a desired outcome~\cite{dumas2013fundamentals,Weske19}, such as having a purchased good available. Van de Ven and Poole refer to these as teleological processes~\cite{van1995explaining}. The analysis focus of BPM is often to determine the start and end states of a process, as well as the sequence of activities that define the progress from the start to the desired goal~\cite[Ch.5]{dumas2013fundamentals}. Typically, many cases run in concurrency according to the same process specification. One of our research partners processes 60 million orders per year. Understanding the differences between these cases offers insights that inform process improvement.

Until now, it has gone largely unnoticed that the focus on teleological processes builds on subtle assumptions that hinder the application of BPM techniques to other categories of processes. One important category of such processes are lifecycle processes~\cite{van1995explaining}. An example of a \textit{lifecycle process} is a patient treated in an elderly care facility. This lifecycle process is better described by trying to keep the patient happy and healthy than by a specific end or outcome. To maintain this status stable, different \textit{vitalizing processes} are executed on a regular or irregular basis that are all targeted at the patient as a focal \textit{entity}. While these vitalizing processes could be analyzed using BPM techniques, it has to be kept in mind that each instance influences the state of the entity during its evolution, as captured by relevant \textit{vital signs}. Clearly, it would be inappropriate to compare the effect of the first with the 20th iteration of chemotherapy.

In this paper, we address this research gap by introducing the notions of a \emph{lifecycle process} and corresponding \emph{vitalizing processes}. With these 
novel categories of processes, we describe processes that are different from typical business processes, such as purchase-to-pay or lead-to-order processes. Instead, these vitalizing processes describe the work performed to prevent an entity from deteriorating or to improve an entity constantly. In this way, vitalizing processes define motors of change for the focal entity and its overarching lifecycle process. Our conceptual contribution provides the basis for the development of novel modeling and analysis techniques that capture the connection between lifecycle and vitalizing processes appropriately.

The remainder of the paper is structured as follows: \autoref{sec:motivation} conceptualizes lifecycle processes and corresponding vitalizing processes for concrete use cases and relates them to prior BPM research. 
\autoref{sec:modeling} introduces a conceptual model that describes the relationships between lifecycle and vitalizing processes.
\autoref{sec:analysis} discusses related work and identifies a set of requirements for the integrated analysis of lifecycle and vitalizing processes, as well as relevant analysis techniques from neighboring fields.
Finally, \autoref{sec:conclusion} summarizes our work and points to future research.

\newcounter{framecnt}
\newenvironment{scenario}[1]
    {\begin{figure}[ht!]
    \refstepcounter{framecnt}
    \renewcommand{\theHfigure}{cont.\arabic{framecnt}}
    \begin{mdframed}[roundcorner=10pt]
    \textbf{\centerline{Box \arabic{framecnt} --- Scenario ``#1''}}
    \smallbreak

    \begin{scriptsize}
    }
    {\end{scriptsize}
    \end{mdframed}
     \end{figure}
    }

\section{Motivation}
\label{sec:motivation} 
This section provides an analysis of lifecycle and corresponding vitalizing processes, and how state-of-the-art BPM techniques are unable to support many analytical questions related to them. First, \autoref{sec:motivation_definition} gives a definition and several examples of lifecycles and vitalizing processes. Next, \autoref{sec:motivation_problems} highlights analytical questions that cannot be answered properly by state-of-the-art BPM techniques.

\subsection{What are Lifecycle and Vitalizing Processes}
\label{sec:motivation_definition}
Lifecycle processes are different from classical business processes. Van de Ven and Poole describe business processes as \emph{teleological}, i.e. directed towards a goal, characterized by planning and organizational problem-solving, with an envisioned end state that is reached by purposeful cooperation~\cite{van1995explaining}. In contrast, \emph{lifecycle processes} exhibit organic growth or decay, driven by a pre-configured program or natural rules. Events progress along a sequence of prescribed stages that are often irreversible~\cite{van1995explaining}.

\begin{figure}[!htb]
    \centering
    \includegraphics[width=.8\textwidth]{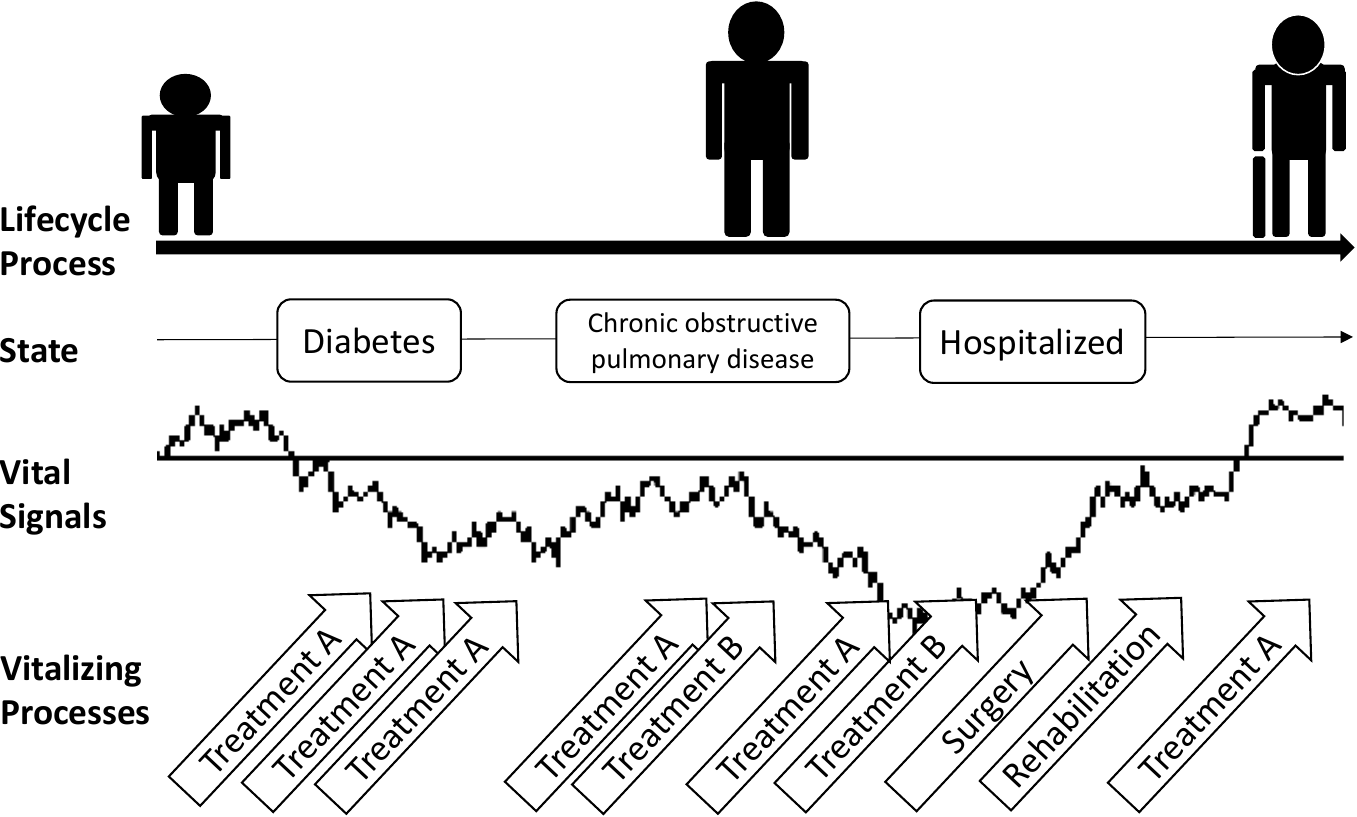}
    \caption{Lifecycle Processes, Vitalizing Processes, and Life Signs related to a Patient as a Focal Entity. }
    \label{fig:concept}
\end{figure}

The essential concepts of lifecycle processes are illustrated in Fig.~\ref{fig:concept}. We define a \emph{lifecycle process} with reference to a focal entity, related events, and vital signs:

\begin{definition}
    A lifecycle process is the sequence of events and observations of vital signs associated with a focal entity over the entire period or a relevant sub-period of its existence.
\end{definition}

With respect to the lifecycle process of a person, we observe the characteristics described above. This lifecycle process lets us expect some sort of organic growth and later decay, driven by the rules of our natural existence, partially in an irreversible way. However, we are not left to the natural rules of growth and decay alone. Various industries, engineering, and research domains develop processes to keep an entity in a desired condition.
Examples of repetitive process structures that stabilize the conditions of an entity are treatment processes to handle the health conditions of a patient, maintenance processes that keep an engine operational, or fertilizing processes that foster the capacity of a field to grow crops. 
Here, we refer to such processes as vitalizing processes, defined as follows:
\begin{definition}
    A vitalizing process is a continuous work process that aims to foster or stabilize the condition of at least one entity with a natural tendency to stagnate or deteriorate. Here, the condition of the entity is described through one or more vital signs that provide a proxy for the current state of the entity. Vitalizing processes have the goal of keeping the entity in the desired condition.
\end{definition}

\noindent We describe three examples of lifecycle and corresponding vitalizing processes. 
\vspace{-4pt}
\begin{enumerate}
    \item \textbf{Chronic Disease Treatment} is a healthcare process focused on one patient as an \textit{entity}~\cite{munoz2022process}. A patient visit to a medical specialist is one of several \textit{vitalizing processes} to ensure stable health. \textit{Vital signs} like blood sugar level or \textit{events} having contracted pneumonia require actions to be taken. 
    
    The \textit{lifecycle process} is the development of the patient over time.
    A \textit{practical goal} here is to stabilize the patient's condition.
    A \textit{modeling goal} is to describe which conditions should be treated with which actions.  An \textit{analysis goal} is to understand which actions are most effective for which condition.

    \item \textbf{Farming} is a process that has been performed by humans since millennia~\cite{dupuis2022predicting}. A field is an \textit{entity} managed to extract a certain crop over an undefined number of seasons. Activities performed by the farmer, such as watering or fertilizing, are \textit{vitalizing processes}, which are motors to ensure optimal conditions for crop growth. The type and amount of actions a farmer performs on the land depend on the state of the soil. \textit{Vital signs} like the level of nitrate, or \textit{events} like a flooding of the field, require certain actions to be taken.
    
    The \textit{lifecycle process} is the development of the field over time.
    A \textit{practical goal} in this context is to improve the condition of growing crops on the field.
    A \textit{modeling goal} is to identify relevant steps and when they should happen in the season, potentially considering environmental conditions.
    An \textit{analysis goal} is to identify which process steps are most beneficial for future seasons, potentially considering crop order.

    \item \textbf{Continuous integration} is a process of enhancing the functionality of a software system as an \textit{entity}. Different \textit{vitalizing processes} serve quality assurance, such as testing, issue management, or bug fixing, together with cycles of incremental extensions according to agile principles~\cite{beck2001manifesto}. \textit{Vital signs} are numbers of fixed bugs or reported issues; \textit{events} like vulnerability reports trigger hotfixes to the software system.
    
    The \textit{lifecycle process} is the development of the software system over time.
    A \textit{practical goal} in this context is to add new features to the system.
    A \textit{modeling goal} is to identify relevant development and quality assurance steps and when they should happen.
    An \textit{analysis goal} is to assess how certain activities, such as reducing technical debt, lead to changes in the productivity of sprints.
\end{enumerate}

\subsection{Challenges for Classical BPM Analysis Techniques}
\label{sec:motivation_problems}
Applying classical BPM analysis techniques for modeling and analyzing the described processes has several issues. A key question for applying many techniques is how to define a case. This comes with commitments to a conceptual window of analysis. In general, the notion of a case can be anchored at the level of the lifecycle process and one specific type of vitalizing process.

First, we can select a specific vitalizing process. Indeed, in the context of chronic disease treatment, this is often one visit to a medical expert~\cite{remy2020event}. In this way, we can analyze how the specific activities performed during a visit to a medical expert impacts health. This selection, however, abstracts from the medical history of the patient and neglects potential interaction with other vitalizing processes, like a new medication given some visits ago.  
Second, we can select the overarching lifecycle process. This might be appropriate for chronic cases, since continuous treatment can grant the affected patient a stable life~\cite{Pauwels2001} and avoid episodes of exacerbation~\cite{Wedzicha2007}. In this way, it allows us to consider the whole history of the patient~\cite{remy2020event}. A challenge is, however, that we have to commit to a specific time window of observation. Also, it might be a challenge to integrate distributed event data from potentially concurrent vitalizing processes in a complexity-controlling way. 
Third, any commitment to a case notion comes with the challenge of integrating analysis methods that focus on events with time series of vital signs.

The above described challenges lead to the question of which requirements must be considered for capturing and analyzing lifecycle and vitalizing processes. Before we analyze our the requirements we first need a conceptual model that describes lifecycle processes in more details.

\section{Modeling Lifecycle and Vitalizing Processes}
\label{sec:modeling}

In this section, we propose a conceptual model for lifecycle processes, vitalizing processes, and their relationships. \autoref{fig:conceptual_model} presents this model as a class diagram. We first discuss the type level, then the instance level. 
At the type level, we distinguish the entity type and its lifecycle process, potentially multiple motor types and multiple vitalizing processes:

\begin{figure}[!htb]
    \centering
\includegraphics[width=\textwidth]{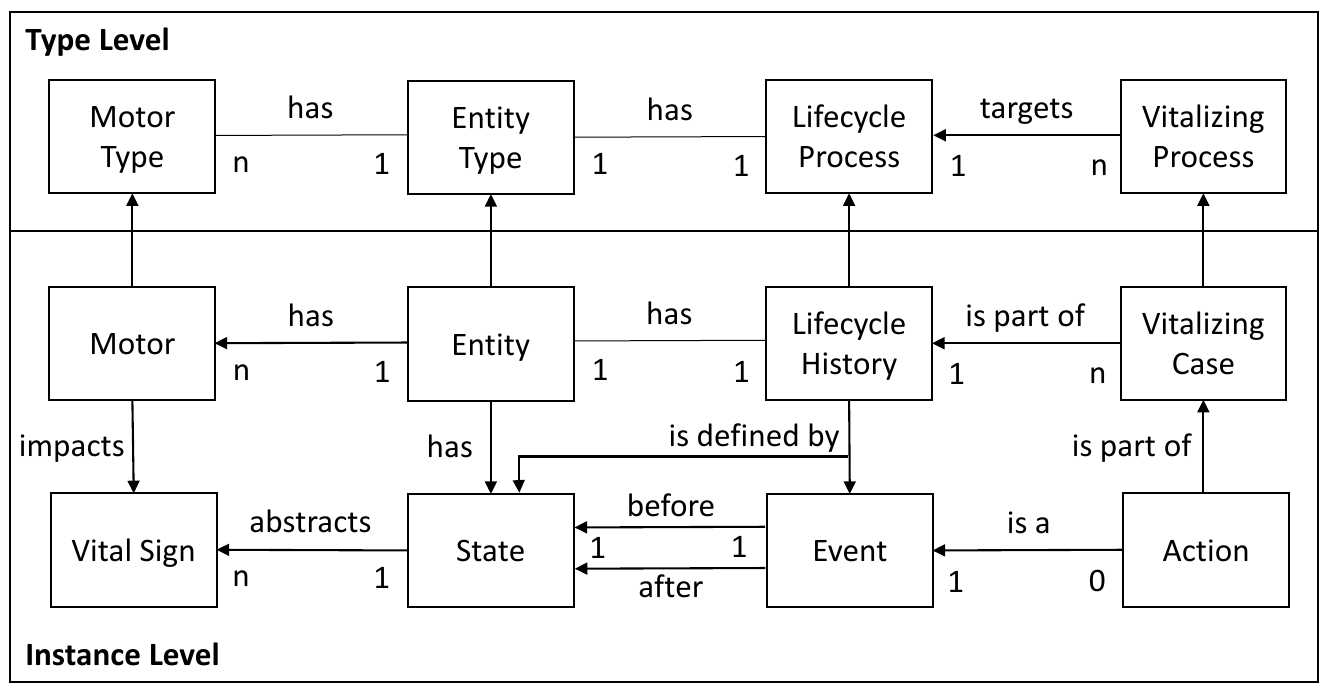}
    \caption{Conceptual Modeling of Vitalizing Business Processes.}
    \label{fig:conceptual_model}
\end{figure}

\mypar{Entity Type} An entity type is a category of entities that share common properties. An entity type has a lifecycle process and several motor types. \textit{Patient} is an entity type. 

\mypar{Lifecycle Process} Each entity type has a lifecycle process that captures its typical evolution. Multiple vitalizing processes can contribute to this evolution. A patient has a typical lifecycle.

\mypar{Motor Type} Each entity type has several motor types that can stabilize or potentially drive its growth or decay. Motor types represent the rules of nature to which an entity type is subject. Examples are the immune reaction of a patient as a stabilizing motor of the current state, puberty as a progressing motor towards a higher level of capabilities, and diseases as deteriorating motors.

\mypar{Vitalizing Process} A vitalizing process is an active intervention targeting the lifecycle process of an entity type. A therapy that can serve as a vitalizing process for a patient. Vitalizing processes typically strengthen stabilizing or progressing motor types or work against the effects of deteriorating motors.

\noindent At the instance level, we distinguish instantiations of the concepts at the type level and more fine-granular temporal concepts.

\mypar{Entity} Each entity instantiates an entity type. It exists in time and space.

\mypar{Motor} A motor is a natural trend that moves the entity towards a certain state. An entity can have multiple motors. A patient can be infected with COVID-19, which has a (potentially only temporary) deteriorating effect on the patient. 

\mypar{Vital Sign} An entity has vital signs. These are measurements of different types at a given point in time, i.e., a patient has a body temperature value of 38.5 degrees on Christmas Eve. 

\mypar{State} An entity has a state. A state can be defined as an abstraction of vital signs. The patient with the mentioned temperature is in a state of fever. 

\mypar{Lifecycle History} An entity has a lifecycle history. It is defined as the collection of all present and past states and events that relate to the entity. A patient has a lifecycle history of medical records.

\mypar{Event} Events define the lifecycle history of an entity. They represent state and vital sign changes of the entity itself, as well as relevant state changes in the environment of the entity. A patient becomes infected with Covid-19.

\mypar{Vitalizing Case} Vitalizing cases are sequences of action that follow the specification of a vitalizing process. A prescription for a fever blocker and its application for a specific patient is a vitalizing case with several regular intakes of pills.

\mypar{Action} An action is a specific type of event where an actor actively changes some state relevant to the entity's lifecycle. For instance, a medical specialist takes a blood sample from a patient.

\section{Analysis of Lifecycle and Vitalizing Processes}
\label{sec:analysis}
This section discusses the requirements for analysis techniques to obtain insights from lifecycle and vitalizing processes. Therefore, \autoref{sec:sota} assesses related work from BPM research, partly supporting the lifecycle process analysis. Then, we identify a deduced set of requirements from the use cases and related work for the analysis of lifecycle and vitalizing processes in \autoref{sec:requirements} Finally, we discuss how techniques from neighboring research areas might fill gaps and provide an overall discussion on what requirements are supported or not by the previously presented related work in~\autoref{sec:discussion_tech}.

\subsection{Related BPM Analysis Techniques}
\label{sec:sota}

There are three main streams of BPM research that relate to lifecycle and vitalizing processes:
\begin{inparaenum}[\itshape i)]
\item approaches that address continuous cyclic structures;
\item data-centric approaches; and 
\item time-series related approaches.
\end{inparaenum}

In the stream of research that tackles \emph{continuous cyclic processes}, Combi et al.~\cite{combi2017methodological} present a proposal for modeling chronic patient treatment with existing BPMN notation elements. The authors highlight the need for a repeating process to stabilize a \gls{copd} patient and capture it as a loop subprocess. The process can also be interrupted in case of unforeseen needs. Additionally, in case of exacerbation, further activities can be executed in parallel to it. This leads to a complex, multi-level process model. 
Strutzenberger et al.~\cite{strutzenberger2021bpmn} extend BPMN to model continuous processes, for instance, beer brewing, by regularly adding ingredients to a brewing process. In their work, continuous processes are characterized by a closed loop, having steady inlet and outlet flows as well as temporally stable conditions. This research highlights that activities in continuous processes must be executed for a particular time span to provide high-end quality. For instance, the deterioration of the reactor over time implies lower-quality beer, but the brewing ingredients and steps remain the same. 
Finally, the issue of repeating process behavior also exists in software engineering when the development process is modelled~\cite{DBLP:journals/infsof/PillatOAC15,moyano2022uses}.
Overarching methodologies with repeating behavior, such as SCRUM, have been modeled using BPMN~\cite{Zaouali2016Agile}. 

In the stream of \emph{data-centric} approaches, the entity, its associated data and characteristics, and its changing states is in the center of process modeling, analysis and automation~\cite{steinau2019dalec}. Such approaches provide means to store information on entities structurally as artifacts, objects, or tuples~\cite{steinau2019dalec}. For example, the artifact-centric approach~\cite{hull2011business} considers an entity's information model, including the attributes, and an entity's lifecycle describing the order of its allowed states. %
Nevertheless, data-centric approaches provide no explicit means to represent the repetitive behavior of cases. 

In the stream of \emph{time-series-related approaches} in the BPM area, several works exists that combine time-series data and event data or process models. 
Dunk et al.~\cite{DBLP:conf/caise/DunklRGF14} consider combining process event logs with time-series data. The objective is to use the time series associated with the activities of the process in order to understand better and explain decision points in the process.
For process analysis, De Smedt et al.~\cite{DBLP:journals/dke/SmedtYPWM23} use time series data to capture behavior drift for process mining discovery.
Lam et al.~\cite{DBLP:journals/eswa/LamIL09} focus on modeling business processes using a custom activity model. This allows to identify empirically inefficiencies and ineffective loops that require intervention. Time series are then used to predict the effects of these interventions.
Finally, in order to been able to handle time-series data, Fonger et al.~\cite{DBLP:conf/caise/FongerAK23} propose to use process models to describe time series data. 
Despite first ideas can be observed, no comprehensive approach exists to relate vital signs, often in the form of time-series data, with the life cycle states of an entity and the events of vitalizing processes.

\subsection{Requirement for Analysis Techniques}
\label{sec:requirements}
The data of lifecycle and vitalizing processes exhibits characteristics that are different from classical business processes. Here, we describe requirements that appropriate analysis techniques have to address. The requirements R1 to R5 are retrieved from the issues addressed by related work, and R6 to R7 are derived from the phenomena of  an entity discussed in the use cases of the previous section.

\mypar{R1: Absence of specific end state}
\textit{We observe} that classical business processes are defined with a clear direction towards a desired end~\cite{Weske19,hull2011business}. Therefore, how to reach efficiently and accurately this end is a pillar for many BPM techniques. 
The \textit{problem} is that lifecycle processes often do not target one specific desired outcome~\cite{strutzenberger2021bpmn}. Rather, the objective of many vitalizing processes is stabilizing or fostering certain conditions of an entity. For example, in healthcare, the death of a patient is avoided by the medical specialists as a factual end. %
Accordingly, we \textit{require} techniques for lifecycle processes that are able to analyze processes that lack a particular end state. Techniques for vitalizing processes must be able to analyze how their different runs impact the entity and, additionally, enable a comparison of them.

\mypar{R2: Recurring behaviour}
\textit{We observe} that classical BPM techniques assume a teleological process from a defined start to a desired end~\cite{Weske19}. Often, this idea is already encoded in the name of the business process, as with order-to-cash, procure-to-pay, and lead-to-close processes~\cite{dumas2013fundamentals}. Recurring behavior can occur within a process execution and needs special attention in the representation, e.g., the iteration workflow patterns~\cite{russell2005workflow}. Recurring behavior is often classified as waste in the process analysis~\cite{dumas2013fundamentals}. 
The \textit{problem} is that vitalizing processes are often enacted repeatedly over the lifecycle of an entity~\cite{combi2017methodological}. This is not a matter of waste,  but a key feature.
Accordingly, we \textit{require} analysis techniques for lifecycle processes that are able to consider recurring behavior. Furthermore, they must be able to incorporate the idea that vitalizing processes might differ, overlap, and depend on each other.

\mypar{R3: Temporal Regularity of Processes}
\textit{We observe} that classical business processes are often expected to be completed as fast as possible~\cite{dumas2013fundamentals}. That is why many BPM methods are focused on remaining time~\cite{verenich2019survey} or waiting time prediction~\cite{senderovich2014queue}. 
The \textit{problem} is that lifecycle processes are often expected to endure as long as possible.
A short cycle time is indeed not desired.
Vitalizing processes are often naturally scheduled at a specific time~\cite{combi2017methodological}. Consider a farming process where the calendar dictates certain activities to be performed. Therefore, the remaining time of a vitalizing process is often known and not a goal for optimization.
Accordingly, we \textit{require} techniques for vitalizing processes that are able to consider the temporal regularity of processes beyond cycle time. %

\mypar{R4: Handling Multi-facet Data}
\textit{We observe} that event logs can only capture changes of attribute values by representing them as new events~\cite{gunther2014xes}. %
This leads to a \textit{problem} that lifecycle processes of an entity, such as a patient, are characterized by their continuously changing vital signs~\cite{DBLP:conf/caise/DunklRGF14}, for instance, to determine if vitalizing actions need to be taken. This might be actively captured, e.g., by a smartwatch with different sensors.
We \textit{require} measures to analyze vital signs over time at varying observation interval density.
While the action on an entity might be represented in events, the information of vital signs is best represented as time series data~\cite{farshchi2018metric}. Thus, we \textit{require} techniques for lifecycle and vitalizing processes to handle multi-facet data consisting of different types.

\mypar{R5: Lifecycle of an Entity}
\textit{We observe} that classical BPM techniques focus on the quality of the process output of each process case in isolation~\cite{dumas2013fundamentals}. 
This is a \textit{problem} because vitalizing processes are executed to archive a stabilization or progression of one or multiple entities. The development over the whole lifecycle process is often more relevant than the results of a single execution of a vitalizing process.
Therefore, we \textit{require} process analysis techniques for lifecycle processes that capture changes in the entity and the trend of its development. %

\mypar{R6: State of an Entity}
\textit{We observe} that classical BPM assumes that each case of a business process has an independent existence. They are executed without reference to each other~\cite{russell2005workflow}. %
This is a \textit{problem} because a vitalizing case needs information about the current state of the entity and the effects of previous vitalizing cases. Thus, a the current state of the entity must be accessible for the vitalizing case.
Accordingly, we
\textit{require} to be able to specify the relevant states and to track them. The lifecycle process and the vitalizing process must be able to read \emph{vital signs} that serve as a proxy for the current state of the entity.

\mypar{R7: Motor as Natural Trend of an Entity}
\textit{We observe} that classical BPM does not consider the natural trend (i.e., the pre-configured program of natural rules) of an entity or its environment. %
This is a \textit{problem} because lifecycle processes are naturally driven by motors associated with the entity. A patient might, for instance, have a disease as a deteriorating motor, which sometimes can be stabilized but not always be stopped. %
Comparing the actual change of the entity with its expected natural trend might give insights into the effectiveness of a vitalizing process.
Accordingly, we \textit{require} means to capture the natural trend of an entity, including the expected effects on the vital signs describing that entity.

\subsection{Other related Techniques and Discussion of Coverage}
\label{sec:discussion_tech}
Based on our requirements analysis, we want to discuss how existing areas of research can contribute to the development of new analysis techniques for lifecycle-related processes. We identified three areas of research that provide us with essential components for future analysis techniques: (i) time series analysis, (ii) social sequence analysis, and (iii) mortality analysis.

\mypar{Time Series Analysis}
The research area of time series~\cite{esling2012time} is concerned with extracting useful information, such as patterns, from time series data. Typical analysis tasks in this area include forecasting time series, clustering similar time series, segmenting a time series, and anomaly detection. However, the most relevant is the area of \emph{motif discovery}~\cite{torkamani2017survey}, identifying frequent unknown patterns with time series. Future techniques for lifecycle-related processes can build on motif discovery and try to link discovered motifs from the vital signs with events or states from the event/state sequences. Furthermore, time series techniques could be applied to vital signs to provide helpful information about the vital signs that could be used to inform actions with vitalizing processes. One specific technique in this context is change point detection. This technique detects the point in time when change occurs~\cite{gama2014survey}, making it an integral part of concept drift analysis. Overall, it can be said that existing techniques from time series analysis provide a substantial component for lifecycle-related processes by helping to utilize hidden information from vital signs. 

\mypar{Social Sequence Analysis}
Social sequence analysis~\cite{abbott1995sequence,cornwell2015social} is concerned with tracing social phenomena over time. The input data is typically defined as a sequence of states~\cite{gabadinho2011analyzing}. Critical design decisions are related to the granularity of the time intervals at which state sequences are captured. Typical analysis tasks for state sequences include describing stochastic patterns, analyzing homogeneity and stationarity, optimal matching, or sequence network construction~\cite{cornwell2015social}. Specifically interesting for analyzing the lifecycle process is stationarity analysis. If a sequence is stationary, then the probability to transition from, e.g., state A to B is the same, no matter in which segment it occurs. %
Understanding such properties might provide indications of recurring behavior, temporal regularity, and the impact of motors on the lifecycle of an entity. Social sequence analysis also offers various techniques for visualizing, sorting, and clustering cohorts of sequences, as well as measures of complexity and turbulence~\cite{gabadinho2011analyzing}. Overall, techniques from social sequence analysis address the requirements of the lifecycle process well because of their assumption that there is no clear start or end. 

\mypar{Mortality Analysis}
Mortality analysis studies the causes of death in populations~\cite{PAHO2018}. This requires various methodologies and approaches to understand mortality rates, life expectancy, and factors influencing death. There are four main types of mortality analysis. First, \textit{descriptive mortality analysis} involves studying death rates, causes of death, and their distribution among different populations, regions, or time periods. Second, \textit{life table analysis} helps analyze mortality rates across different age groups, providing insights into life expectancy and survival probabilities. Third, \textit{cause-specific mortality analysis} examines the causes of death within a population to understand disease burdens and health priorities. Fourth, \textit{f} compares mortality rates across different demographic groups, socio-economic classes, or geographic regions to identify disparities and underlying factors.\\
Mortality analysis has also seen broader applications such as in finance~\cite{Altman2000}, forest research~\cite{hiroshima2014applying}, and even history~\cite{steckel1979slave}. Future lifecycle-related techniques can draw inspiration from mortality analysis in multiple ways. The four types of mortality analysis can be applied to the lifecycle history. To this end, events of lifecycle processes will need to be gathered and analyzed. Consequently, vitalizing processes can be redesigned to counter the discovered issues.

\begin{table}
    \centering
    \begin{tabularx}{\textwidth}{l|X X X X X X X} %
        \toprule
         & R1 & R2 & R3 & R4 & R5 & R6 & R7 \\ %
         \midrule
        Combi et al.~\cite{combi2017methodological} & \cmark & \cmark & \cmark & \xmark & \xmark & \xmark & \xmark \\
        Strutzenberger et al.~\cite{strutzenberger2021bpmn}  & \cmark & \cmark & \cmark & \xmark & \xmark & \xmark & \xmark \\
        Hull et al.~\cite{hull2011business} & \xmark & \xmark & \xmark &  \xmark & \cmark & \cmark &\xmark \\
        Fonger et al.~\cite{DBLP:conf/caise/FongerAK23} & \xmark & \xmark & \cmark & (\cmark) & \xmark & \cmark & \xmark \\
        Dunk et al.~\cite{DBLP:conf/caise/DunklRGF14}  & \xmark & \xmark & \xmark & \cmark &\xmark & \xmark& \xmark \\
        Farschi et al.~\cite{farshchi2018metric} & \cmark & \cmark & \xmark & \cmark & \xmark & \xmark & \xmark \\
        \midrule
        Time Series Analysis & \cmark & \cmark & \xmark & \xmark & \cmark & \cmark & \cmark \\
        Social Sequence Analysis & \cmark & \cmark & \cmark & \xmark & \cmark & \cmark & \cmark \\
        Mortality Analysis & \cmark & \xmark & \xmark & \xmark & \cmark & \cmark & \cmark \\
        \bottomrule
    \end{tabularx}
    \caption{Overview of requirements for lifecycle-related process analysis techniques and how they are supported by existing techniques. \cmark: Is supported; (\cmark): Is partially supported \xmark: Is not supported.}
    \label{tab:req_bpm_techniques}
\end{table}

\mypar{Overall Comparision} In \autoref{tab:req_bpm_techniques}, we highlight what requirements of lifecycle and vitalizing processes are fulfilled by existing techniques and neighboring research areas. We can observe that most BPM techniques are only focused on specific sub-problems of lifecycle-related business processes. As an example, the idea of repeating behavior, as presented in vitalizing cases, is discussed in several papers~\cite{combi2017methodological,strutzenberger2021bpmn}. 
Similarly, several papers~\cite{DBLP:conf/caise/DunklRGF14,DBLP:conf/caise/FongerAK23} try to integrate time series, as in the form of vital signs, into BPM techniques. 
In future work, it will be possible to integrate the results from such research into techniques that address lifecycle and vitalizing processes.

Furthermore, we can observe that many requirements for lifecycle and vitalizing processes have been addressed in other research fields such as \textit{time-series}, \textit{social-sequence} and \textit{mortality} analysis (rows 6--8). Notably, these research fields are usually not concerned with the data fusion necessary to address lifecycle and vitalizing business processes. Nonetheless, we can highlight that they support many of the aspects necessary for analysis technique of lifecycle-related processes and therefore can serve as valuable building blocks for novel approaches.

Overall, we can argue that many approaches specialized on specific aspects of lifecycle and vitalizing processes exist. Therefore, we believe that this phenomenon can be addressed by combining existing BPM techniques and knowledge from other areas of research.

\section{Conclusion}
\label{sec:conclusion}

In this paper, we showed that traditional BPM techniques predominantly address teleological business processes with defined start and end states, e.g., purchase-to-pay processes. This research highlighted the existence of \emph{lifecycle process} of a focal entity, such as a patient, displaying either a natural increase or diminution. They are supported by \emph{vitalizing processes} that foster or stabilize the lifecycle process.
We provided a conceptual model with the key ingredients of a lifecycle process and a data model. 
We discussed the requirements of analysis techniques for lifecycle and vitalizing processes and showed that previous research work only covered parts of them. We have come to the conclusion that many specialized aspects of lifecycle and vitalizing processes are already supported by existing research. This existing research can serve as a foundation for analysis techniques targeted towards lifecycle processes. 

This paper is centered on identifying and conceptualizing the problem space without offering direct solutions. We envision this groundwork as a catalyst for a new line of research within BPM. In the future, we plan to develop appropriate analysis techniques for lifecycle and their vitalizing processes and plan to apply them to real-world data.  Furthermore, an important next step for the research lies in the collection of data from the lifecycle and vitalizing business processes.

\section*{Acknowledgements}
This work was supported by the German Federal Ministry of Education and Research (BMBF), grant number 16DII133 (Weizenbaum-Institute). We thank Julian Theis for fruitful discussions about lifecycle processes in industrial settings.

 \bibliographystyle{splncs04}
 \bibliography{mybib}
\end{document}